\documentclass[prb,aps,epsfig,twocolumn,showpacs,preprintnumbers,superscriptaddress,float,amsmath,amssymb]{revtex4-1}
\usepackage{graphicx}
\usepackage{dcolumn}
\usepackage{bm}

\begin{document}

\title{Anomalous and spin Hall effects in hcp cobalt from GGA + $U$ calculations}

\author{Jen-Chuan Tung}
\affiliation{Graduate Institute of Applied Physics, National Chengchi University, Taipei 11605, Taiwan}
\author{Huei-Ru Fuh}
\affiliation{Graduate Institute of Applied Physics, National Chengchi University, Taipei 11605, Taiwan}
\author{Guang-Yu Guo}\email{gyguo@phys.ntu.edu.tw}
\affiliation{Graduate Institute of Applied Physics, National Chengchi University, Taipei 11605, Taiwan}
\affiliation{Department of Physics and Center for Theoretical Sciences, Nation Taiwan University, Taipei 10617, Taiwan}

\date{\today}

\begin{abstract}
We have calculated the intrinsic anomalous and spin Hall conductivities, spin and orbital magnetic moments
and also spin polarization of Hall currents 
in hexagonal cobalt 
within the density functional theory 
with the generalized gradient approximation (GGA) plus on-site Coulomb interaction (GGA+$U$). 
The accurate all-electron full-potential linearized augmented plane wave method is used. 
We find that the anomalous Hall conductivity (AHC) ($\sigma_{xy}^A$) and orbital magnetic moment ($m_o$) of cobalt 
with the magnetization being along the $c$-axis ($m//c$) calculated in the GGA+$U$ scheme with $U = 1.6$ eV and $J = 0.9$ eV
agree rather well with the corresponding experimental values while, in contrast,  
the $\sigma_{xy}^A$ and $m_o$ from the GGA calculations are significantly smaller than the measured ones.
This suggests that moderate $U = 1.6$ eV and $J = 0.9$ eV are appropriate for Co metals.
The calculated AHC and spin Hall conductivity (SHC) ($\sigma^S$) 
are highly anisotropic, and the ratio of the Hall conductivity for $m//c$ to
that for the magnetization being in-plane ($m//ab$) from the GGA+$U$ calculations,
is $\sim$16.0 for the AHC and $\sim$6.0 for the SHC.
For $m//c$, the spin-up and spin-down Hall currents are found to flow in the same direction.
The magnitude of the calculated Hall current spin polarization ($P^H$) is large and $P^H = -0.61$.
Remarkably, for $m//ab$, the spin-up and spin-down Hall currents are predicted to flow in the opposite
directions. This indicates that the Hall current contains both spin-polarized charge
current and pure spin current, resulting in the magnitude of the Hall current spin polarization ($P^H = -1.59$)
being larger than 1.0. 
 
\end{abstract}

\pacs{71.15.Mb, 71.70.Ej, 72.15.Gd, 72.25.Ba}

\maketitle

\section{Introduction}

Spin-dependent electronic transports have attracted intensive attention recently 
because they not only are interesting from the view-point of fundamental physics but also have 
fascinating technological applications. 
Anomalous Hall effect (AHE), discovered in 1881 by Hall\cite{Hal81},
is an archetypal spin-related transport phenomena and hence has received renewed 
interests in recent years.\cite{Nag10}
The ordinary Hall effect in nonmagnetic conductors is driven by the Lorentz 
force due to the applied magnetic field while the AHE depends on magnetization in
ferromagnetic materials\cite{Chi80}. The dipole magnetic field produced by the magnetization 
is too small to explain the observed AHE in ferromagnets, and thus it has been 
realized for long time that the AHE
must be caused by the spin-orbit interaction (SOI). 
Therefore, several competing SOI-based mechanisms have been proposed. Extrinsic mechanisms of
skew scattering\cite{Smit55} and side jump\cite{Berger70} refer to the modified impurity scattering 
induced by the SOI, and hence should in principle depend on the nature and density of the impurities. 
Surprisingly, extrinsic side jump contribution to the anomalous Hall conductivity (AHC) turns
out to be fairly independent of either scattering strength or defect density.\cite{Berger70,Nag10}
Another mechanism arises from the anomalous velocity of the Bloch electrons caused by the 
SOI, discovered by Karplus and Luttinger\cite{Karplus29}, and is thus of intrinsic nature. 
Interestingly, this intrinsic mechanism has recently been reinterpreted in terms of the Berry curvature of the occupied
Bloch states.\cite{Cha96,Jun02,Xia10} Furthermore, recent first-principles studies based on
the Berry phase formalism showed that the intrinsic AHE is important in various materials.\cite{Yao04,Zen06}
In particular, in itinerant ferromagnets such as Fe, the intrinsic AHC
given by first-principles density functional calculations with the generalized gradient
approximation (GGA)\cite{Yao04} has been found to agree rather well with the experimental
AHC\cite{Dhe67,Tia09}. 

Nonetheless, the physical origin of the AHE in cobalt and some other ferromagnets are still not fully understood. 
Recent GGA calculations predicted an intrinsic AHC of $\sim$480 S/cm in hcp Co with the magnetization 
along the $c$-axis, being smaller than the corresponding experimental value by about 40 \%.\cite{Roman09}
Density functional calculations with either the local density approximation (LDA) or GGA have been 
rather successful in describing many physical
properties such as crystal structure, elastic constant and spin magnetic moment,
of itinerant ferromagnets Fe, Co and Ni (see, e.g., Ref. \onlinecite{Guo00} and references therein).
However, LDA and GGA calculations often fail in describing relativistic SOI-induced phenomena
in these itinerant magnets.  For example, the theoretical values of orbital magnetic
moment account for only about 50 \% of the measured ones in Fe and Co\cite{Daa90,Guo91} and the calculated
magnetocrystalline anisotropy energy of hcp Co is even wrong in sign\cite{Daa90}. This failure of the LDA/GGA 
is generally attributed to their incomplete treatment of the 3$d$ electron-electron correlation in these systems.
Several theoretical methods that go beyond the LDA and GGA, such as the orbital-polarization correction (OPC)\cite{Bro85}
and LDA/GGA plus on-site Coulomb interaction $U$ (LDA/GGA+$U$)\cite{Ani91,Czy94,Lie95} schemes,
have been developed for better description of the SOI-induced phenomena in magnetic solids.
Indeed, the orbital-polarization correction has been found to bring the calculated orbital
moments in itinerant magnets such as Fe and Co in good agreement with experiments\cite{Guo97}.
It has also been demonstrated that the correct easy axes and the magnitudes of
the magnetocrystalline anisotropy energy of Co\cite{Try95} as well as Fe and Ni\cite{Yan01} 
can be obtained within either the OPC\cite{Try95} or LDA+$U$ scheme\cite{Yan01}.
Furthermore, recent GGA+$U$ calculations\cite{Fuh11} showed that including 
appropriate on-site Coulomb interaction $U$ could reduce the theoretical AHC by half,
bringing the calculated intrinsic AHC in good agreement with the latest experiments.

The AHE is closely connected to the spin Hall effect (SHE) \cite{Dya71,Mur03,Sin04,Kat04}, which refers
to the transverse spin current generation in nonmagnetic materials by an applied electric field. 
In analogous to the AHE, the mechanisms for the SHE can be separated into intrinsic and extrinsic types.
Both intrinsic AHE and SHE are caused by the opposite anomalous velocities experienced by 
the spin-up and spin-down electrons when they move through the relativistic energy bands in solids
under the electric field. In nonmagnetic materials where the numbers of the spin-up and
spin-down electrons are equal, these opposite transverse currents would give rise to a pure 
spin current. On the other hand, in ferromagnets, where an unbalance of spin-up and spin-down
electrons exists, the same process would result in a charge current.
The SHE has attracted intensive current interests both experimentally and theoretically
since the theoretical proposal for its intrinsic mechanism in semiconductors\cite{Mur03,Sin04},
because it would enable us to control spins without magnetic field or magnetic materials.
Indeed, first-principles density functional calculations on the intrinsic spin Hall conductivity
(SHC) have recently carried out for both nonmagnetic semiconductors\cite{Guo05,Yao05} and 
metals\cite{Yao05,Guo08,Guo09,Fre10}. And the calculated SHCs\cite{Guo08,Guo09} are found
to agree well with the corresponding experimental values in many metals such as Al\cite{Val06}, 
Pt\cite{Kim07,Mor11}, Au and Pd\cite{Mos10}, suggesting that the intrinsic SHE is dominant 
in the high resistivity regime in pure metals. Nonetheless, recent experimental\cite{Sek08,Nii11} and 
theoretical\cite{Guo09a,Gra10,Gu10,Low11,Fer11} studies showed that
certain impurities could also give rise to large SHE in impure metals. 
However, the SHE in ferromagnets remains unexplored.

In this work, we therefore carry out GGA+$U$ calculations of intrinsic anomalous and spin Hall
conductivities and relativistic band structure as well as spin and orbital magnetic moments
of hcp Co. The primary objective of this work is to better understand the AHE 
and its anisotropy in hcp Co. In particular, we would examine how the theoretical AHC and related
physical quantities would be affected when the on-site
Coulomb interaction is taken into account in the GGA+$U$ scheme. Secondly, we would study
the intrinsic SHE and its anisotropy in hcp Co. Among the elemental ferromagnetic transition
metals, cobalt is promising for future spintronic devices because of its large
magnetocrystalline anisotropy energy and highly anisotropic AHC.
A knowledge of the spin-polarization of the intrinsic Hall current 
is important for the spintronic applications. Therefore, another objective
of this work is to evaluate the degree of spin-polarization of the charge Hall current
by using the calculated AHC and SHC. 
This paper is organized as follows. In the next section, 
we briefly describe how the intrinsic Hall conductivities are calculated within
the linear-response Kubo formalism as well as the numerical method and computational details
used here. In Sec. III, we present the calculated intrinsic Hall
conductivities, magnetic moments, relativistic band structure and also
the spin polarization of charge Hall currents. We then compare our
results with available experiments and also previous calculations. 
In Sec. IV, we summarize the main conclusions drawn from the present work.
 
\section{Theory and Computational Method}

An intrinsic Hall conductivity of a solid can be evaluated by using the Kubo formalism\cite{Marder00}.
The intrinsic Hall effect comes from the static limit ($\omega$=0) of the corresponding off-diagonal element of the optical conductivity\cite{Marder00,Guo05,Fuh11}.
Following the procedure for the calculations of the magneto-optical conductivity\cite{Guo95}, we first calculate
the imaginary part of the off-diagonal elements of the optical conductivity

\begin{eqnarray}\label{Imopti}
\sigma^{(2)}_{ij}(\omega)=-\frac{\pi e}{\omega V_c}\sum_{\bf k}\sum_{n\neq n^{'}}&(f_{{\bf k}n}-f_{{\bf k}n^{'}})Im[\langle {\bf k}n|j_{i}|{\bf k}n^{'}\rangle\notag\\
                                             &\times \langle{\bf k}n|v_{j}|{{\bf k}n^{'}}\rangle]\delta (\hbar \omega - \epsilon_{n^{'}n}) \notag\\
 \end{eqnarray}
where $V_{c}$ is the unit cell volume, $\hbar\omega$ is the photon energy, $|{\bf k} n\rangle$ is the $n$th Bloch state with crystal
momentum ${\bf k}$, $v_{j}$ is the $j$ component of the velocity operator, and $\epsilon_{n^{'}n}=\epsilon_{{\bf k}n^{'}}-\epsilon_{{\bf k}n}$. 
Here $j_i$ is the $i$ component of the current operator which is $-ev_i$ and $\frac{\hbar}{4}\{\sigma_k,v_i\}$ for 
the anomalous and spin Hall effects, respectively.\cite{Guo05} And also, $k\ne i\ne j$.
We then obtain the real part from the imaginary part by a Kramers-Kroning transformation

\begin{equation}\label{KKtrans}
\sigma^{(1)}_{ij}(\omega)=\frac{2}{\pi}{\bf P}\int^{\infty}_{0}{\bf d}\omega^{'}\frac{\omega^{'}\sigma^{(2)}_{ij}(\omega^{'})}
 {\omega^{'2}-\omega^{2}}
 \end{equation}
where {\bf P} denotes the principal value of the integral. The intrinsic Hall conductivity $\sigma^{H}_{ij}$ is the static
limit of the off-diagonal element of the optical conductivity $\sigma^{(1)}_{ij}(\omega=0)$. We notice that the anomalous Hall conductivity of
bcc Fe {\cite{Yao04,Guo95} and also the spin Hall conductivity of fcc Pt\cite{Guo08,Gra11} calculated in this way are in good agreement with that calculated directly by
accounting for the Berry phase correction to the group velocity.

Since all the intrinsic Hall effects are caused by the spin-orbit interaction, theoretical calculations must be based on a relativistic
band theory. Here the relativistic band structure of hcp Co is calculated using the highly accurate all-electron
full-potential linearized augmented plane wave (FLAPW) method, as implemented
in the WIEN2K code\cite{wien2k02}. The self-consistent electronic structure calculations are based on the GGA for the
exchange correlation potential\cite{Perdew96}. To further take $d$-electron correlation into account, we introduce on-site Coulomb
interaction $U$ in the GGA+$U$ approach\cite{Ani91,Lie95}. The double counting correction scheme proposed by Czyzyk
and Sawatzky\cite{Czy94} is used here. $U=1.6$ eV and $J=0.9$ eV, which were found to give the correct orbital magnetic moment 
for hcp Co, are used (see Sec. III below). Further calculations with a larger $U$ value of 1.9 eV and 2.5 eV are also performed
to examine how the variation of $U$ may affect the calculated anomalous and spin Hall conductivities.

The experimental lattice constants\cite{Marder00} $a$=2.51 (\AA) and $c$=4.07 (\AA) are used here. The muffin-tin sphere radius
($R_{mt}$) used is 2.2 a.u. The wave function, charge density, and potential were expanded in terms of the spherical harmonics inside
the muffin-tin spheres and the cutoff angular moment ($L_{max}$) used is 10, 6 and 6, respectively. The wave function outside the muffin-tin sphere
was expanded in terms of the augmented plane waves (APWs) and a large number of APWs (about 80 APWs per atom, i. e., the maximum
size of the crystal momentum $K_{max}=8/R_{mt}$) were included in the present calculations. The improved tetrahedron method is used for the
Brillouin-zone (BZ) integration\cite{Blochl94}. To obtain accurate ground state charge density as well as
spin and orbital magnetic moments, a fine 69$\times$69$\times$37 grid of 176157 $k$-points  in the first BZ was used.

\begin{figure}
\includegraphics[width=8cm]{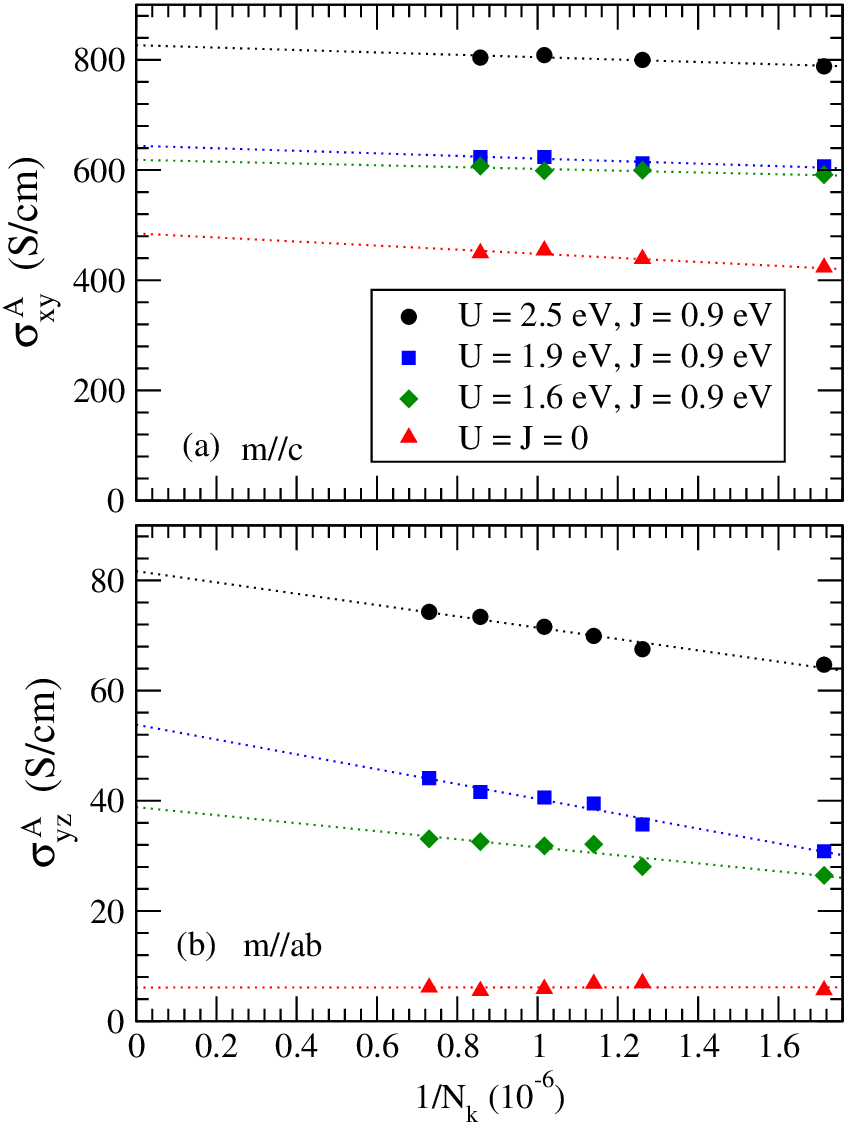}\\
\caption{(color online) Calculated anomalous Hall conductivity  $\sigma_{ij}^{A}$  with and without 
on-site Coulomb interaction $U$, for  
the magnetization being either along the $c$-axis (a) or in the $a$-$b$ plane (b), as a function 
of the inverse of the number of $k$-points $N_k$. 
}
\label{Fig1}
\end{figure}

\section{Results and discussion}

\subsection{Anomalous and spin Hall conductivities}

\begin{table}
\caption{Calculated anomalous [$\sigma_{ij}^{A}$ (S/cm)] and spin [$\sigma_{ij}^{S}$ ($\hbar$S/ecm)] 
Hall conductivities as well as 
spin [$m_{s}$ ($\mu_{B}$/atom)] and orbital [$m_{o}$ ($\mu_{B}$/atom)] magnetic moments of hcp Co 
with magnetization being
along either the $c$-axis or in the $a$-$b$ plane. 
The on-site Coulomb interaction $U$ = 1.6, 1.9 and 2.5 eV are used and $J$ = 0.9 eV is kept the same.  
The estimated experimental value of the scattering independent Hall conductivity
$\sigma_{ij}^{ind}$
as well as previous theoretical intrinsic $\sigma_{ij}^{A}$ and extrinsic side jump conductivity
$\sigma_{ij}^{SJ}$ are also listed for comparison.
Note that the estimated $\sigma_{ij}^{ind}$ should be compared with the theoretical 
$\sigma_{ij}^{A}$ plus theoretical $\sigma_{ij}^{SJ}$.}
\begin{ruledtabular}
\begin{tabular}{cccccc}
                   & GGA   & GGA+$U$      & GGA+$U$      & GGA+$U$      & Expt.         \\
                   &       &  $U=1.6$ eV  & $U=1.9$ eV   & $U=2.5$ eV   &               \\ \hline
                   &  &\multicolumn{3}{c}{{\bf m}$\parallel c$}&              \\
 $\sigma_{xy}^{A}$ & 484 &      618   &      643   &      827   &                     \\
                   & 481\footnotemark[1]  &            &         &         &    \\
$\sigma_{xy}^{SJ}$ & 217\footnotemark[2]       &            &         &         &    \\
$\sigma_{xy}^{ind}$ &                           &            &         &         &813\footnotemark[3]\\
$\sigma_{xy}^{S}$  & -117                     & -188       &   -195  &   -318  &    \\
 $m_{s}$           &   1.63  &  1.69  &  1.71   &  1.75   &  1.52\footnotemark[4] \\
 $m_{o}$           &   0.081  &  0.161 &  0.193   &  0.287  &  0.14\footnotemark[4] \\
                   &  &\multicolumn{3}{c}{{\bf m}$\parallel ab$}&     	          \\
 $\sigma_{yz}^{A}$ &   6.1     &     38.9        &      53.8  &   81.7    &                        \\
                   &   116\footnotemark[1]   &        &      &      &               \\
$\sigma_{yz}^{SJ}$ &   -30\footnotemark[2]     &        &      &      &               \\
$\sigma_{yz}^{ind}$ &                           &            &         &         &150\footnotemark[3]\\
$\sigma_{yz}^{S}$  & -14.9                     & -30.9      &   -32.9 &   -29.6 &    \\
 $m_{s}$           &  1.63   &   1.71          &     1.71       &     1.75   &             	\\
 $m_{o}$           &   0.077 &   0.136         &     0.175       &     0.268    &             	\\

\end{tabular}
\footnotetext[1]{GGA calculations (Ref. \onlinecite{Roman09}).}
\footnotetext[2]{GGA calculations (Ref. \onlinecite{Wei11}).}
\footnotetext[3]{Estimated experimental values (Ref. \onlinecite{Roman09})}
\footnotetext[4]{Experimental values (Ref. \onlinecite{Bon86})}
\end{ruledtabular}\label{table1}
\end{table}

Calculated anomalous and spin Hall conductivities 
with and without on-site Coulomb interaction are listed in Table \ref{table1}. Like the calculation 
of the magnetocrystalline anisotropy energy of bulk magnets\cite{Daa90,Guo91}, 
a very fine $k$-point mesh is needed for 
the anomalous and spin Hall conductivity calculations. Therefore, we perform the Hall conductivity 
calculations using several very fine $k$-point meshes with the finest $k$-point mesh being 137$ \times$137$\times$73. 
The calculated anomalous Hall conductivity $\sigma_{ij}^{A}$ is plotted as a function of the inverse of 
the number ($N_k$) of $k$ points in the first Brillouin zone in Fig. 1. The calculated values of $\sigma_{ij}^{A}$ 
are fitted to a first-order polynomial to get the converged theoretical $\sigma_{ij}^{A}$ (i. e., the extrapolated 
value of $\sigma_{ij}^{A}$ at $N_k$=$\infty$) (see Fig. 1).  
The theoretical anomalous Hall conductivity $\sigma_{ij}^{A}$ and also spin Hall conductivity $\sigma_{ij}^{S}$ 
obtained in this manner are listed in Table \ref{table1}.

First of all, we notice that taking the on-site Coulomb interaction into account
via the GGA+$U$ scheme affects the calculated AHC  $\sigma_{ij}^{A}$ and SHC $\sigma_{ij}^{S}$ significantly
(see Table \ref{table1} and Fig. \ref{Fig1}).
In particular, when the magnetization is along the $c$-axis ($m//c$), the theoretical $\sigma_{xy}^{A}$ from 
the present GGA calculations is 484 S/cm. When the on-site Coulomb interaction is taken into account
in the GGA+$U$ scheme, the calculated AHC  $\sigma_{xy}^{A}$  increases significantly with increasing $U$.  
For example, at $U =1.6$ eV, the theoretical  $\sigma_{xy}^{A}$ becomes 618 S/cm, being increased by $\sim$30 \%. 
If $U$ is further increased to 2.5 eV, the calculated $\sigma_{xy}^{A}$ is 827 S/cm.
Similar behavior is found for the $\sigma_{yz}^{A}$ with the magnetization being in the $a$-$b$ plane ($m//ab$) 
(see Table \ref{table1} and Fig. \ref{Fig1}). Secondly, as reported before in Ref. \onlinecite{Roman09},
the intrinsic AHC in hcp Co is highly anisotropic. The AHC for $m//c$ is more than 10 times larger than 
that for $m//ab$ (see Table \ref{table1}). 

The theoretical SHCs of nonmagnetic hcp metals were recently reported\cite{Fre10}. Interestingly, we find 
that the calculated $\sigma_{xy}^{S}$ for hcp Co is larger than that of any 3$d$ hcp transition metals 
(see Table \ref{table1} and Ref. \onlinecite{Fre10}). Furthermore, hcp Zn was found to have the largest 
anisotropy among the nonmagnetic metals studied in Ref. \onlinecite{Fre10}. The calculated SHC of hcp Co 
is rather anisotropic. The SHC for $m//c$ is more than 6 times larger than that for $m//ab$ (Table \ref{table1}), 
and hence the anisotropy is nearly the same as that of hcp Zn. 
The calculated $\sigma_{yz}^{S}$ is largest among all the 3$d$ hcp transition metals.

The AHC $\sigma_{xy}^{A}$ from the present calculations is in good agreement with previous theoretical calculations\cite{Roman09},
whilst, in contrast, the AHC $\sigma_{yz}^{A}$ presented here is significantly smaller than that reported in 
Ref. \onlinecite{Roman09} (see Table \ref{table1}). 
The precise reasons for this discernable discrepancy 
in $\sigma_{yz}^{A}$ between the present and previous calculations\cite{Roman09} are not known.
Note that $\sigma_{yz}^{A}$ is one order of magnitude smaller than $\sigma_{xy}^{A}$ (see Fig. 1). 
On the one hand, the discrepancy could be attributed to the fact that much more $k$-points could be included 
in the BZ integration in Ref. \onlinecite{Roman09} 
where much more efficient Wannier function interpolation technique was used. 
Nevertheless, Fig. 1 indicates that the theoretical $\sigma_{yz}^{A}$ values presented in Table I 
are converged with respect to the number of $k$-points used.
In Ref. \onlinecite{Yao04}, it was pointed out that the calculated $\sigma_{ij}^{A}$ could be
very sensitive to the fine details of the band structure especially near the Fermi level. 
As Figs. 2 and 3 indicate, the band structure of hcp Co is rather complex near the Fermi level. 
On the other hand, therefore, the discrepancy could also be caused by the fact that
the band structure of hcp Co generated by the Wannier functions might not precisely 
reproduce the one calculated by using plane wave pseudopotential method in Ref. \onlinecite{Roman09}.
Another reason could be the fact that the dipole matrix elements for Eq. (1) are calculated
using the full APW wave functions in the present calculations while they were calculated using 
Wannier functions in Ref. \onlinecite{Roman09}. 

In Ref. \onlinecite{Roman09}, the experimental scattering-independent AHC was derived from 
formula $\sigma_{xy}^{ind} \approx \rho_{xy}^A/\rho_{xx}^2$ using the old measured Hall and longitudinal
resistivities from Refs. \onlinecite{Vol61} and \onlinecite{Mas66}, respectively, and is 813 S/cm
(see Table \ref{table1}). This experimental AHC value should contain both the intrinsic AHC $\sigma_{xy}^{A}$ and
extrinsic scattering-independent side jump contribution $\sigma_{xy}^{SJ}$. The theoretical side jump $\sigma_{xy}^{SJ}$ from
the recent {\it ab initio} calculations\cite{Wei11} is 217 S/cm. Therefore, for $U = 1.6$ eV,
the theoretical $\sigma_{xy}^{A}$ plus $\sigma_{xy}^{SJ} = 217$ S/cm is almost in perfect agreement 
with the estimated experimental value of $\sigma_{xy}^{ind}$. 
Nevertheless, it should be cautioned that this perfect agreement could be fortuitous because in the estimation\cite{Roman09}
of the experimental $\sigma_{xy}^{ind}$ the Hall and longitudinal resistivities were taken from two 
separated old measurements\cite{Vol61,Mas66}. Furthermore, the theoretical side jump contributions 
listed in Table \ref{table1} were calculated by considering only the short-range disorder in the weak
scattering limit\cite{Wei11} and hence may not reflect the total side jump mechanism.
As will be reported in the next subsection, the 
calculated spin and orbital magnetic moments for $U = 1.6$ eV are also in good agreement with the experiments, 
indicating that the appropriate $U$ value for Co metals should be about 1.6 eV. 
The experimental AHC $\sigma_{yz}^{ind}$ was also estimated using more complicated formula 
$\sigma_{yz}^{ind} = \sigma_{xy}^{A}(\rho_{yz}^A/\rho_{xy}^A)(\rho_{xx}/\rho_{zz})$ in Ref. \onlinecite{Roman09},
and is 150 S/cm. The reported theoretical value\cite{Wei11} of the side jump $\sigma_{yz}^{SJ}$
is -30 S/cm. Consequently, the theoretical value (8.9 S/cm at $U = 1.6$ eV) of $\sigma_{yz}^{A}$ 
plus $\sigma_{yz}^{SJ}$ is much too small in comparison with the estimated experimental $\sigma_{yz}^{ind}$.
Again, given the fact that $\rho_{yz}^A/\rho_{xy}^A$ and $\rho_{xx}/\rho_{zz}$ were taken
from two separated old measurements\cite{Vol61,Mas66}, such a comparison of the estimated experimental value of
$\sigma_{ij}^{ind}$ with the calculated $\sigma_{ij}^{A}$ plus $\sigma_{ij}^{SJ}$ might not be very meaningful. 
Unfortunately, recent measurements of anomalous Hall and longitudinal resistivities on a same
speciman were conducted only on polycrystalline Co metals\cite{Koe05,Shi10}, and the deduced experimental
values of the anomalous Hall conductivity $\sigma_{ij}^{ind}$ are $\sim$200 S/cm (Ref. \onlinecite{Koe05})
and $\sim$500 S/cm (Ref. \onlinecite{Shi10}). Clearly, further simultaneous 
measurements of both longitudinal and Hall resistivities on the well-characterized thin films such as 
that on bcc Fe reported in Ref. \onlinecite{Tia09} are needed to extract reliable 
anomalous Hall conductivities for hcp Co, especially, $\sigma_{yz}^{A}$.

\subsection{Spin and orbital magnetic moments}

In addition to the AHC and SHC, including the SOI in the fully relativistic calculations would also 
give rise to nonzero orbital magnetic moments in the magnetic solids concerned.
The calculated spin and orbital magnetic moments are listed in Table \ref{table1}.
First of all, Table \ref{table1} shows that both the spin and orbital magnetic moments
increases monotonically with increasing on-site Coulomb interaction $U$. However, this 
increase in the orbital magnetic moment is much more significant than that of the spin magnetic moment.
As mentioned earlier, due to its incomplete treatment of the 3$d$ electron-electron correlation,
the GGA calculations give an orbital moment which is only half of the corresponding experimental
value\cite{Bon86}. In contrast, the theoretical orbital moment from the GGA+$U$ calculations with $U=1.6$ eV
is almost doubled and becomes in rather good agreement with the experimental value (see the results for m//c in 
Table \ref{table1}). When the $U$ is further increased, the orbital moment becomes
much larger than the experimental value. This suggests that $U = 1.6$ eV would be appropriate
for Co metals. Interestingly, as mentioned before, the orbital magnetic moments from the OPC
calculations also agree well with the experiments\cite{Guo97,Try95}.  

Secondly, Table \ref{table1} indicates that the orbital magnetic moment depends significantly
on the magnetization orientation, whereas the spin magnetic moment is not. Furthermore, this 
magnetocrystalline anisotropy in the orbital magnetic moment increases pronouncedly when
the on-site Coulomb interaction is taken into account. For example, the orbital moment anisotropy
from the GGA calculations is about 5.0 \% and becomes $\sim$16.0 \% in the GGA+$U$ calculations
with $U = 1.6$ eV.  

\begin{figure}
\includegraphics[width=8cm]{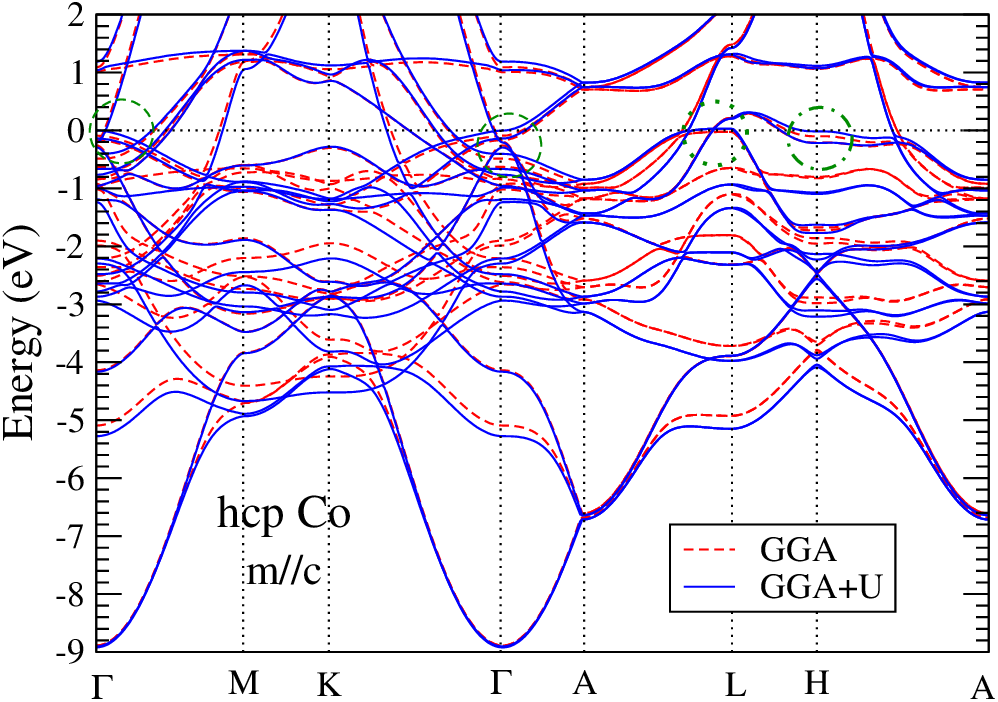}\\
\caption{(color online) Relativistic band structures calculated with (solid lines) 
and without (dashed lines) on-site Coulomb interaction $U =1.6$ eV.
The magnetization is along the $c$-axis. The Fermi level is shifted to 0 eV.}
\label{Fig2}
\end{figure}

\subsection{Effects of on-site Coulomb interaction on band structure}

To investigate how the on-site electron-electron correlation affects the electronic band structure and AHE in cobalt, 
we plot the relativistic energy bands along the high-symmetric lines in the Brillouin zone calculated 
with and without on-site Coulomb interaction for the magnetization being along
the $c$-axis in Fig. \ref {Fig2} and in the $a$-$b$ plane in Fig. \ref{Fig3}. 
The relativistic band structure of hcp Co has been studied before by several researchers
by using different band structure calculation methods (see, e.g., Refs. \onlinecite{Daa90} and \onlinecite{Roman09} and references
therein). The present relativistic GGA band structures (the red dashed curves in Figs. 2, 3, and 4)
are nearly identical to that reported in Refs. \onlinecite{Daa90} and \onlinecite{Roman09}.
The relativistic band structure may be regarded as
the results of a superposition of the corresponding scalar-relativistic spin majority and minority band structures with many accidental band
crossings lifted by the SOI. These SOI-induced splittings in cobalt are, in general, much smaller than the exchange splittings
(see Fig. \ref{Fig2} and Fig \ref{Fig3}), and thus can be treated as a perturbation.  

Nonetheless, it is the inclusion of the SOI that makes the band structure depend on the
magnetization orientation and this magnetization dependence of the band structure results in
a strong anisotropy in such SOI-induced quantities as anomalous and spin Hall conductivities. 
An example of clear changes in the band structure due to the rotation of the magnetization
from the $c$-axis to an in-plane direction is the energy bands just below the Fermi level ($E_F$) in the
vicinity of the H point (as highlighted by the dash-dotted green circle in Fig. 2 and Fig. 4a).
These energy bands are significantly spin-orbit split when the magnetization
is along the $c$-axis but they remain degenerate when the magnetization is in-plane
(see Figs. 2, 3, and 4). Furthermore, the SOI-induced splittings for the bands near the $E_F$
at the $\Gamma$ point when the magnetization is in-plane, are nearly two times larger 
than that of the magnetization along the $c$-axis. These differences in the band structure
near the $E_F$ due to different magnetization orientations thus give rise to 
a pronounced anisotropy in the SOI-induced phenomena such as the AHC, SHC and orbital magnetic moment. 

A pronounced change in the band structure due to the inclusion of the on-site Coulomb interaction is that all the 
$d$-dominant bands are lowered in energy by about 0.2 eV relative to the $E_F$ 
(see Figs. 2, 3, and 4). This is clearly seen in the energy window between -5.5 and -0.5 eV.
On the other hand, the energy bands near and above the $E_F$ are much less affected 
when the on-site Coulomb interaction is taken into account, as shown in Figs. 2, 3, and 4.
In particular, for the in-plane magnetization, the energy bands in this energy range
remain almost unchanged upon the inclusion of the on-site Coulomb interaction (see Fig. 3 and
Fig. 4d). With respect to the AHC for the magnetization along the $c$-axis, the most 
significant change due to $U$ is perhaps that the energy band in the vicinity of the 
$\Gamma$, L and H symmetry points (as highlighted by the green circles in Fig. 2 and Fig. 4a)
which is just below the $E_F$ in the GGA
calculations, is now pushed onto the $E_F$ in the GGA+$U$ calculations. As a result,
the AHC $\sigma_{xy}^A$ is increased by about 30 \% upon including the on-site Coulomb interaction
$U = 1.6$ eV.

\begin{figure}
\includegraphics[width=8cm]{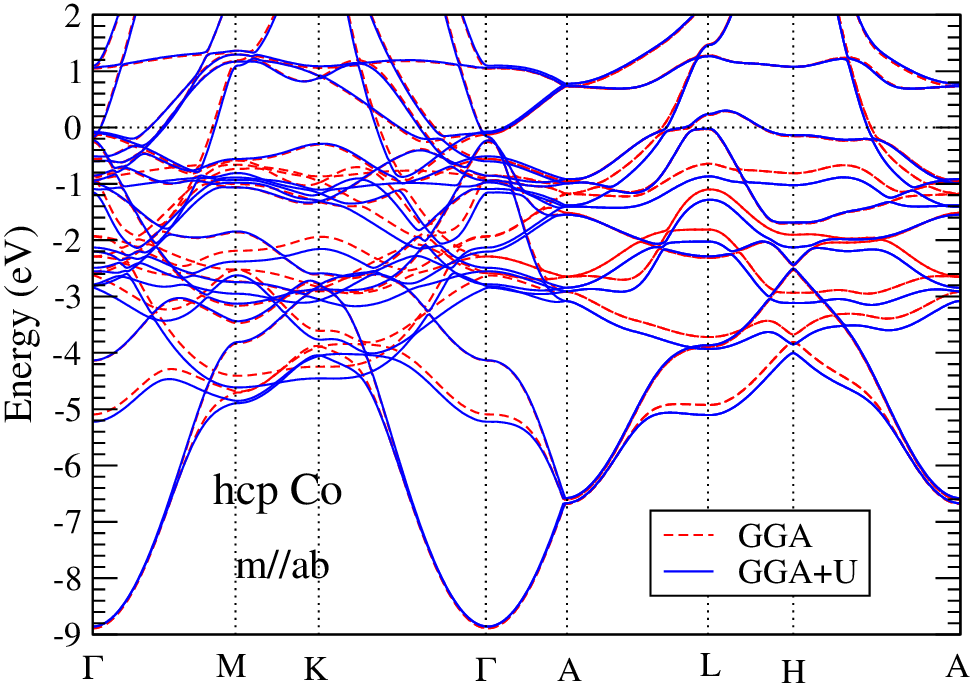}\\
\caption{(color online) Relativistic band structures calculated with (solid lines) and 
without (dashed lines) on-site Coulomb interaction $U =1.6$ eV.
The magnetization is in the $a$-$b$ plane. The Fermi level is shifted to 0 eV.}
\label{Fig3}
\end{figure}

In Fig. \ref{Fig4}, we display the calculated anomalous Hall conductivity ($\sigma_{ij}^{A}$) 
and the number of valence electrons per Co atom($n_e$/atom) as a function of the $E_F$, 
together with the relativistic band structures without and with on-site 
Coulomb interaction using $U=1.6$ eV and $J= 0.9$ eV. The fine $k$ point mesh of 130$\times$130$\times$69 
is used. Fig. \ref{Fig4}b shows that in general, the $\sigma_{xy}^{A}$ calculated with and 
without the on-site Coulomb interaction are rather similar. The most significant difference is 
that the $\sigma_{xy}^{A}$ at the true Fermi level (0 eV) from the GGA+$U$ calculation is 30 \% larger than 
that obtained from the GGA calculation. As mentioned in the preceding paragraph,
this enhancement of the $\sigma_{xy}^{A}$ could be attributed to the fact that the inclusion of
the on-site Coulomb interaction pushes the energy band near the $\Gamma$, L and H points 
to just above the Fermi level (Fig. 2 and Fig. 4a). Fig. \ref{Fig4}e indicates that for $m//ab$, the 
two $\sigma_{yz}^{A}$ spectra calculated with and without the on-site Coulomb repulsion 
are nearly identical, except their magnitudes differ slightly here and there.   

Fig. 4b also shows that the calculated $\sigma_{xy}^{A}$ are quite flat in the vicinity of the true Fermi level. 
As the Fermi level is artificially lowered from about -0.3 eV to -0.71  eV ($n_e$/atom = $\sim$7.95), 
the $\sigma_{xy}^{A}$ increases steadily and eventually peaks at -0.71  eV with a maximum value of
$\sim$2600 S/cm. When the Fermi level is further lowered, the $\sigma_{xy}^{A}$ first oscilates and
then starts to decrease sharply at -1.2 eV, and eventually, it changes its sign at -1.3 eV. 
If the Fermi level is artificially raised above 0 eV, the $\sigma_{xy}^{A}$ decreases gradually and
then changes to the negative value at $\sim$0.5 eV. The $\sigma_{xy}^{A}$ reaches a maximum magnitude 
(-1900 S/cm) at
$\sim$1.15 eV. Above this energy, the magnitude of the $\sigma_{xy}^{A}$ decreases steadily and becomes
very small above $\sim$ 2.0 eV. Fig. 4e suggests that the calculated $\sigma_{yz}^{A}$ for $m//ab$
is overall similar to the $\sigma_{xy}^{A}$ for $m//c$ (Fig. 4b), especially above -1.5 eV.
In particular, the $\sigma_{yz}^{A}$ spectra also have a negative peak (-800 S/cm) near 1.25 eV and
a positive peak (1500 S/cm) around -0.70 eV. Nonetheless, the peaks in the $\sigma_{yz}^{A}$
are smaller than the corresponding ones in the $\sigma_{xy}^{A}$ spectra. 

\begin{figure}
\includegraphics[width=8cm]{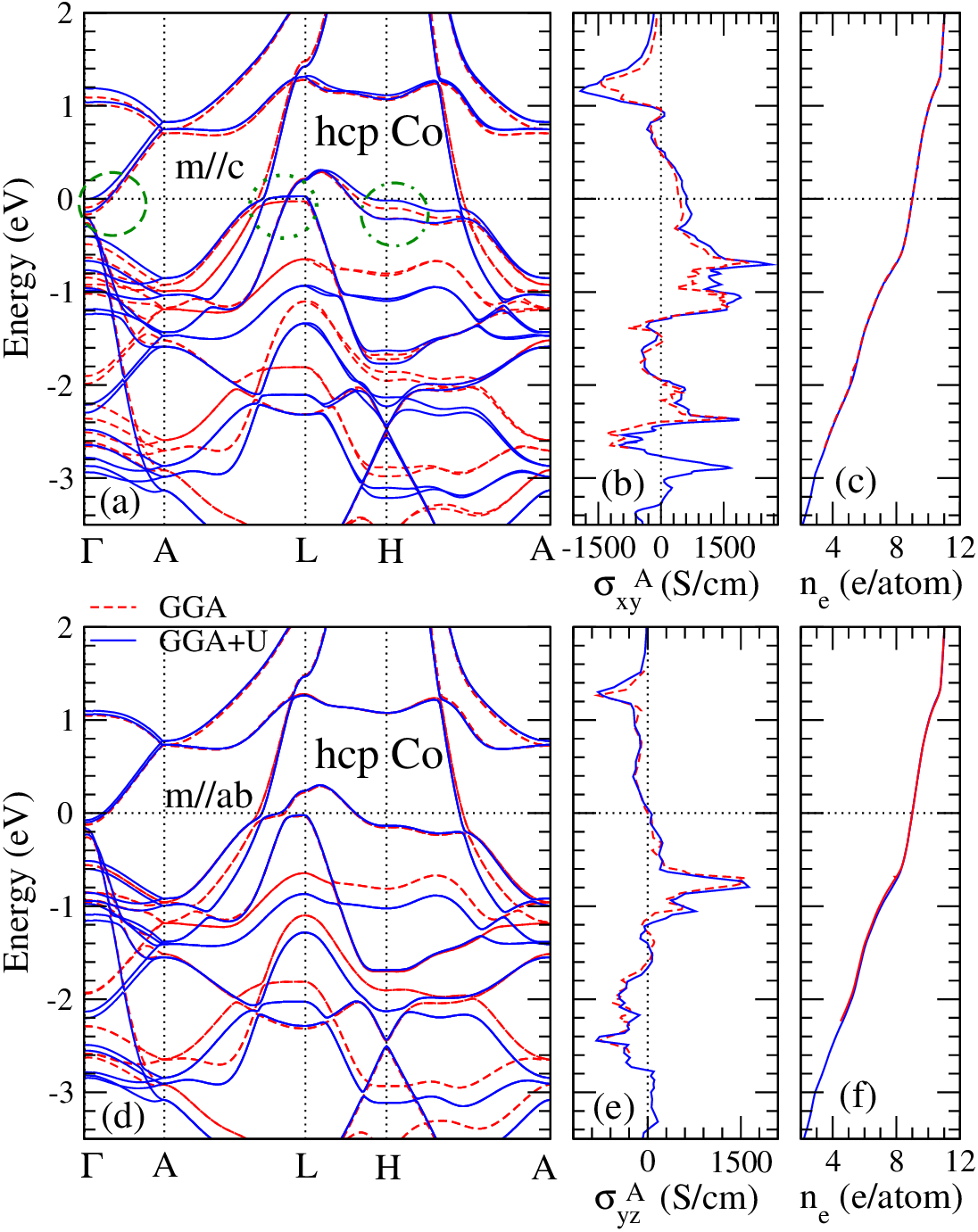}\\
\caption{(color online) Relativistic band structures calculated with (solid lines) and 
without (dashed lines) on-site Coulomb interaction $U =1.6$ eV (a, d),
anomalous Hall conductivity ($\sigma_{ij}^{A}$) (b, e), and  number ($n_e$) of valence electrons per Co atom (c, f).
Upper panels are for the magnetization along the $c$-axis ($m//c$) and lower panels are for the magnetization 
in the $a$-$b$ plane ($m//ab$). The Fermi level is shifted to 0 eV. The green circles in (a) are a guide to
the eyes only.}
\label{Fig4}
\end{figure}

\subsection{Spin polarization of Hall current}

Electric and spin current transports are determined by the characteristics of the band structure near the
Fermi level in the solids concerned. For the spintronic applications, it would be interesting to
examine the extent that the density of states (DOS) near the $E_F$ of the systems considered are spin-polarized. 
This spin-polarization ($P^{DOS}$) of the DOSs near the $E_F$ is defined as 
\begin{equation}
P^{DOS}=\frac{N_{\uparrow}(E_F)-N_{\downarrow}(E_F)}{N_{\uparrow}(E_F)+N_{\downarrow}(E_F)},
\end{equation}
where $N_{\uparrow}(E_F)$ and $N_{\downarrow}(E_F)$ are the spin-up and spin-down DOSs at the $E_F$, 
respectively. $P^{DOS}$ may vary from -1.0 to 1.0. For the half-metallic materials, $P^{DOS}$ is either 
1.0 or -1.0, and the charge current would be fully spin polarized. 
We have calculated DOSs of hcp Co with the magnetization being along the $c$-axis and in the $a$-$b$ plane,
and the calculated spin-decomposed DOSs at the $E_F$ are listed in Table \ref{table2}. The corresponding
spin polarization $P^{DOS}$ defined above for both magnetization directions is also listed in Table \ref{table2}. 
It is clear from Table \ref{table2} that the spin polarization is negative and the degree of spin polarization
is rather high, being $66\sim69$ \%. Furthermore, the spin polarization $P^{DOS}$ is independent of 
magnetization orientation, in strong contrast to the high anisotropy of the calculated anomalous and spin
Hall conductivities (Table \ref{table1}). We note that the current spin polarization of cobalt measured in the 
tunneling experiments\cite{Mon00} and in the Andreev reflection experiments\cite{Str01} is about
42 \% and 45 \%, respectively, being considerably lower than the calculated  $P^{DOS}$.
 
As pointed out by several researchers before (see, e.g., Ref. \onlinecite{Maz99}), the spin polarization
$P^{DOS}$ defined by Eq. (3) is not the spin-polarization of the transport currents measured in the experiments.
The anomalous Hall effect has recently received intensive renewed interest mainly because of its 
close connection with spin transport phenomena\cite{Nag10}. In particular, for spintronic applications,
it could be advantageous to use the Hall current from ferromagnets as a spin-polarized current source, 
instead of the longitudinal current,
because of the topological nature of the intrinsic AHE\cite{Mur03}. Therefore, it would be interesting to know 
the spin polarization of the Hall current.  
To this end, let us define the spin-polarization $P_{ij}^{H}$ of the Hall current as 
\begin{equation}
P_{ij}^{H}=\frac{\sigma_{ij}^{H\uparrow}-\sigma_{ij}^{H\downarrow}}{\sigma_{ij}^{H\uparrow}+\sigma_{ij}^{H\downarrow}},
\end{equation}
where $\sigma_{ij}^{H\uparrow}$ and $\sigma_{ij}^{H\downarrow}$ are the spin-up and spin-down 
Hall conductivities, respectively. Note that this decomposition of the Hall conductivity 
in terms of the simple two current model, works reasonably well only for metals containing light
elements such as 3$d$ transition metals, but may fail for metals containing heavy elements
such as Co$_{1-x}$Pt$_x$ alloys\cite{Low10}. The $\sigma_{ij}^{H\uparrow}$ and $\sigma_{ij}^{H\downarrow}$ can
then be obtained from the calculated AHC $\sigma_{ij}^{A}$ and SHC $\sigma_{ij}^{S}$ via the following relations
\begin{equation}
\sigma_{ij}^{H\uparrow}+\sigma_{ij}^{H\downarrow} = \sigma_{ij}^{A},\hspace{5mm}
\sigma_{ij}^{H\uparrow}-\sigma_{ij}^{H\downarrow} = 2\frac{\hbar}{e}\sigma_{ij}^{S}.
\end{equation}
The $\sigma_{ij}^{H\uparrow}$, $\sigma_{ij}^{H\downarrow}$, and $P_{ij}^{H}$ obtained in this way,
are listed in Table \ref{table2}.
Note that, unlike the spin-polarization of the DOSs and also of the longitudinal charge
currents, the magnitude of the $P_{ij}^{H}$ can be larger than 1.0. And this is because
spin-decomposed Hall currents can go either right (positive) or left (negative). 
In the nonmagnetic materials, the charge Hall current is zero, and hence 
$\sigma_{ij}^{H\uparrow} = -\sigma_{ij}^{H\downarrow}$. In this case, we would then have 
pure spin current, and  $P_{ij}^{H}=\infty$.

Interestingly, Table \ref{table2} shows that, in contrast to $P_{ij}^{DOS}$,
the spin polarization of charge Hall current is highly anisotropic. In particular,
when the magnetization is parallel to the $c$-axis, both spin-up and spin-down Hall
conductivities are positive and hence the magnitude of $P_{ij}^{H}$ is smaller than 1.0.
The magnitude of both $P_{ij}^{H}$ for $U = 0$ and 1.6 eV is rather large, being
0.48 and 0.61, respectively, although it is smaller than the corresponding
$P^{DOS}$ value. In contrast, for the in-plane magnetization with 
$U = 0$ and 1.6 eV, the spin-up and spin-down Hall conductivities have an opposite sign.
Therefore, the magnitude of $P_{ij}^{H}$ is larger than 1.0, being 4.90 and 1.59 for 
$P_{ij}^{H}$ for $U = 0$ and 1.6 eV, respectively. Remarkably, this suggests that the Hall current
could be a good spin-polarized current source. The best materials known for
the spin-polarized current source are pressumably the half-metallic magnets with the
spin-polarization being 1.0 (see, e.g., Ref. \onlinecite{Wan06} and references therein). 
The longitudinal spin-polarized current from a magnetic metal is charge current
while the Hall current could consist of both spin-polarized charge current and pure
spin current, as mentioned before. Pure spin current is dissipationless 
while charge current would consume energy.\cite{Mur03} Therefore, using the Hall current
instead of the longitudinal current as the spin-polarized current source could
also save energy. 

Table \ref{table2} also indicates that $P_{ij}^{H}$
can be significantly affected by the on-site Coulomb interaction.  
For example, upon the inclusion of the on-site Coulomb interaction $U = 1.6$ eV,
the magnitude of  $P_{ij}^{H}$ would be increased by 25 \% when the magnetization is 
along the $c$-axis, but would be decreased by $\sim$ 65 \% when the magnetization is
in the $a$-$b$ plane (see Table \ref{table2}). There has been no report on
the measurement of the spin polarization of the anomalous Hall current.
Therefore, one could be tempted to compare the theoretical $P_{ij}^{H}$ with
the measured spin polarization ($P^L$) of the longitudinal
charge current\cite{Mon00,Str01}. However, the theoretical $P_{yz}^{H}$ 
calculated using $U = 0$ and 1.6 eV is very different from the
measured $P^L$, although the theoretical $P_{xy}^{H}$ is comparable
to the $P^L$ (see Table \ref{table2}). This suggests that the spin polarizations
of the longitudinal and Hall currents are two different things and may have no
correlation. It is hoped that the interesting findings reported in this paper
will stimulate the measurements on the spin polarization of the Hall current
in either hcp Co or other magnetic metals. 

\begin{table}
\caption{Theoretical spin-decomposed Hall conductivities 
[($\sigma_{ij}^{H\uparrow}$ and $\sigma_{ij}^{H\downarrow}$ (S/cm)] and spin-polarization of the
Hall current ($P^{H}$) as well as spin-decomposed densities of states ($N_{\uparrow}$ and $N_{\downarrow}$) 
(states/eV/unit cell) and spin-polarization of the density of states at the
Fermi level ($P^{DOS}$). The on-site Coulomb interaction $U$ = 1.6, 1.9 and 2.5 eV are used, 
with $J$ = 0.9 eV being constant. The measured longitudinal current spin-polarizations ($P^L$)
are also listed for comparison.}
\begin{ruledtabular}
\begin{tabular}{cccccc}
                                  & GGA      & GGA+$U$     & GGA+$U$      & GGA+$U$       & Expt.\\
                                  &          & $U=1.6$ eV  &  $U=1.9$ eV  &  $U=2.5$ eV   &      \\ \hline
                                  &  &\multicolumn{3}{c}{{\bf m}$\parallel c$}&              \\
 $\sigma_{xy}^{A\uparrow}$        &   125   &    121 &   127   &    96   &             \\
 $\sigma_{xy}^{A\downarrow}$      &   359   &    497 &   517   &    732  &             \\
    $P_{xy}^{H}$                  &   -0.48  &     -0.61 &    -0.61   &    -0.77   &         \\
    $N_{\uparrow}$                & 0.316   &  0.281         &   0.291    & 0.257  &  \\
    $N_{\downarrow}$              & 1.485   &  1.509         &   1.504    & 1.400  &  \\
    $P^{DOS}$                     &   -0.66  &     -0.68 &    -0.68   &    -0.69   &         \\
    $P^{L}$                       &          &           &            &            & -0.42\footnotemark[1] \\
                                  &          &           &            &            & -0.45\footnotemark[2] \\
                                  &  &\multicolumn{3}{c}{{\bf m}$\parallel ab$}&         \\
 $\sigma_{yz}^{A\uparrow}$        &   -11.9  &   -11.5   &    -6.0    &   11.3     &               \\
 $\sigma_{yz}^{A\downarrow}$      &   18.0   &    50.4   &    59.8    &   70.5     &               \\
 $P_{yz}^{H}$                     &  -4.90   &   -1.59   &    -1.22   &   -0.72    &               \\
    $N_{\uparrow}$                & 0.302    &  0.293    &   0.290    & 0.253      &   \\
    $N_{\downarrow}$              & 1.491    &  1.539    &   1.554    & 1.626      &   \\
 $P^{DOS}$                        &  -0.66   &   -0.68   &    -0.68   &   -0.69    &               \\

\end{tabular}
\footnotetext[1]{Tunneling experiments (Ref. \onlinecite{Mon00}).}
\footnotetext[2]{Andreev reflection measurements (Ref. \onlinecite{Str01}).}
\end{ruledtabular}\label{table2} 
\end{table}

\section{Conclusions}
We have presented theoretical intrinsic anomalous and spin Hall conductivities, spin and orbital magnetic moments
and also spin polarization of Hall currents in hcp Co with different magnetization orientations
calculated in both the GGA and GGA+$U$ schemes.  The accurate FLAPW method is used.
First of all, we find that the AHC ($\sigma_{xy}^A$) and orbital magnetic moment ($m_o$) of cobalt
with $m//c$ calculated in the GGA+$U$ scheme with $U = 1.6$ eV and $J = 0.9$ eV
are in good agreement with the corresponding experimental values while, in contrast,
the $\sigma_{xy}^A$ and $m_o$ from the GGA calculations are significantly smaller than the measured ones.
This suggests that the on-site Coulomb interaction, though moderate only in Co metals, should be
taken into account properly in order to get SOI-induced properties.
The most obvious effect of including the on-site Coulomb interaction on the band structure is
that all the $d$-dominant bands in the energy window between 0.5 and 5.5 eV
below the $E_F$ are lowered in energy relative to the $E_F$
by about 0.2 eV while the energy bands above the $E_F$ remain nearly unaffected.
In the context of the Hall conductivities, the most
significant change due to $U$ is that the energy band in the vicinity of the
$\Gamma$, L and H symmetry points just below the $E_F$ in the GGA
calculations, is pushed onto the $E_F$ in the GGA+$U$ calculations, resulting
in a $\sim$30 \% increase in the $\sigma_{xy}^A$ and also a $\sim$60 \% increase in the $\sigma_{xy}^S$.
Secondly, we find that the calculated AHC and SHC ($\sigma^S$)
are highly anisotropic, and the ratio of the Hall conductivity for $m//c$ to
that for $m//ab$ from the GGA+$U$ calculations, is $\sim$16.0 for the AHC and $\sim$6.0 for the SHC.
Finally, the calculated Hall current spin polarization ($P^H$) is -0.61 for $m//c$ and -1.59 for $m//ab$.
The theoretical $\sigma^S$ and $P^H$ are also found to be significantly affected by
the inclusion of the on-site Coulomb interaction $U$.
We hope that this work would stimulate simultaneous 
measurements of both longitudinal and Hall resistivities on the well-characterized thin films
to extract reliable experimental anomalous Hall conductivities for hcp Co, especially, $\sigma_{yz}^{A}$.

\section*{Acknowledgments}
The authors gratefully acknowledge supports from the National Science Council and the National Center 
for Theoretical Sciences of Taiwan as well as the Center for Quantum Science and Engineering, 
National Taiwan University (CQSE-10R1004021). They also thank the
National Center for High-performance Computing of Taiwan for computer time and facilities.
Finally, G. Y. Guo thanks Yugui Yao for stimulating discussions on calculating spin Hall
conductivity using the WIEN2k code.


\end{document}